\documentclass[amssymb,prl,twocolumn,notitlepage,longbibliography,superscriptaddress]{revtex4-1}

\renewcommand{\vec}[1]{\mathbf{#1}}

\usepackage[%
colorlinks=true,
urlcolor=blue,
linkcolor=blue,
citecolor=blue
]{hyperref}

\usepackage{color}

\usepackage{amsmath}                          
\usepackage{amsfonts}                         
\usepackage{amssymb}                          
\usepackage{amsthm}                           
\usepackage{mathdots}                         

\usepackage{graphicx}                         
\usepackage{subfigure}                        
\usepackage{overpic}

\begin{document}


\preprint{APS/123-QED}
\title{Quantum-enhanced least-square support vector machine: simplified quantum algorithm and sparse solutions}

\author{Jie Lin}
\altaffiliation{Co-first author}
\affiliation{Henan Key Laboratory of Quantum Information and Cryptography, IEU, Zhengzhou, Henan 450001, China}
\author{Dan-Bo Zhang}
\altaffiliation{Co-first author}
\affiliation{Guangdong Provincial Key Laboratory of Quantum Engineering and Quantum Materials, GPETR Center for Quantum Precision Measurement, SPTE, South China Normal University, Guangzhou 510006, China}
\author{Shuo Zhang }
\author{Xiang Wang }
\author{Tan Li}
\author{Wansu Bao}
\email{bws@qiclab.cn}
\affiliation{Henan Key Laboratory of Quantum Information and Cryptography, IEU, Zhengzhou, Henan 450001, China}

\date{\today}

\begin{abstract}
Quantum algorithms can enhance machine learning in different aspects. Here, we study quantum-enhanced least-square support vector machine (LS-SVM). Firstly, a novel quantum algorithm that uses continuous variable to assist matrix inversion is introduced to simplify the algorithm for quantum LS-SVM, while retaining exponential speed-up. Secondly, we propose a hybrid quantum-classical version for sparse solutions of LS-SVM. By encoding a large dataset into a quantum state, a much smaller transformed dataset can be extracted using quantum matrix toolbox, which is further processed in classical SVM. We also incorporate kernel methods into the above quantum algorithms, which uses both exponential growth Hilbert space of qubits and infinite dimensionality of continuous variable for quantum feature maps. The quantum LS-SVM exploits quantum properties to explore important themes for SVM such as sparsity and kernel methods, and stresses its quantum advantages ranging from speed-up to the potential capacity to solve classically difficult machine learning tasks.
\end{abstract}

\maketitle


\section{\label{sec:level1}Introduction}
Supervised learning aims to establish a mapping between features and labels of data by learning from a dataset~\cite{qiu2016survey,dunjko2018machine,novakovic2016toward,hsu2002comparison}. The learned mapping is further used to classify new data. A support vector machine (SVM) serves as a classifier by learning a hyperplane that separates two classes of data in a feature space, where the feature map can be implicitly generated with kernel methods~\cite{wu2005svm,suykens1999least,scholkopf1999input,sanchez2003advanced}. As a result of spare solutions, the number of samples involved for inference can be small. Due to the flexibility of incorporating kernels and the sparsity of solutions, SVM is powerful and efficient in classifying complicated data and thus has wide real-world applications.\\
\indent Recently, the frontier of machine learning is pushed forward with quantum computing~\cite{gu2012quantum,sarma2019machine,wiebe2012quantum,dunjko2016quantum,lloyd2016quantum,huang2017experimental,lloyd2018quantum}, which utilizes the capacity of quantum computing to manipulate data in a large Hilbert space. By exploiting quantum properties such as superposition and entanglement, machine learning can be enhanced in several aspects, such as quantum speed-up, quantum-enhanced feature maps, better generalization, and so on~\cite{zhang2019physical,bremner2016average,douce2017continuous}.\\
\indent As for the SVM, a quantum algorithm with exponential speed-up has been proposed~\cite{rebentrost2014quantum} and verified with a proof-of-principle experiment~\cite{li2015experimental}. Kernel methods has also been discussed, where feature map is done explicitly by encoding classical data as quantum states~\cite{benedetti2019parameterized, schuld2019quantum,havlivcek2019supervised}. However, the quantum algorithm is designed for a least-square version of SVM, which is lack of sparsity. Also, quantum matrix inversion~\cite{harrow2009quantum,rebentrost2018quantum}, a main ingredient of the algorithm, involves many ancillary qubits and complicated circuits for eigenvalue/singular-value inversion. Stressing those issues can help realize feasible quantum algorithm for support vector machine with less quantum resources in near-term quantum technology.\\
\indent In this paper, we study quantum-enhanced least-square support vector machine, and propose a simplified quantum algorithm as well as sparse solutions. Remarkably, those quantum algorithms are designed deliberately so as to be applicable when considering quantum feature maps. Firstly, we propose a quantum algorithm that assists the matrix inversion with two continuous variables~\cite{lloyd2003hybrid,lau2017quantum,zhang2019realizing,arrazola2018quantum}, which greatly simplifies the algorithm for quantum LS-SVM and reduces the use of ancillary qubits. Secondly, we give sparse solutions for quantum LS-SVM. The large classical dataset is firstly encoded into a quantum state, and then is compressed into a much smaller dataset, while essential information for classification task is kept. This procedure is conducted with quantum matrix toolbox, including quantum principal component analysis (QPCA)~\cite{lloyd2014quantum} and quantum singular-value threshold algorithm (QSVT)~\cite{duan2017quantum,duan2018efficient}. The transformed dataset is further processed using classical LS-SVM. In this regard, it is a hybrid quantum-classical LS-SVM. The quantum enhancement manifests in a manipulation of data with speed-up with quantum matrix toolbox, as well as quantum feature maps that may be classically intractable~\cite{bremner2016average,douce2017continuous,schuld2019quantum,havlivcek2019supervised,harrow2009quantum}. Our investigation of quantum-enhanced LS-SVM thus contributes to important topics in SVM with feasible quantum algorithms that exploits quantum advantages.\\
\indent  The remainder of the paper is organized as follows: We introduce an improved quantum hybrid variables method, give a brief overview of the LS-SVM algorithm in Sec. II, put forward a quantum LS-SVM with hybrid variables (HVQ-SVM) algorithm and propose a hybrid variables algorithm for sparse solution of LS-SVM model (QSLS-SVM) in Sec. III, discuss the regularization problem and advantages of algorithm in Sec. IV,  present our conclusions in Sec. V.\\
\section{\label{sec:level2}The review of SVM, LS-SVM and an improved quantum hybrid variables method}
In this section, we firstly introduce the general notations and the SVM, LS-SVM model~\cite{wu2005svm,suykens1999least}. Next, we present an improved quantum hybrid variables method based on the Ref.\cite{zhang2019realizing,arrazola2018quantum}.
\subsection{Notations}
The each vector $x$ can be defined as $y=(y_{1},y_{2},\cdots,y_{M}),y\in R^{1\times N}$. A set of training data is defined as $\{x_{i},y_{i}\},i=1,2,\cdots,M$, where $x_{i}\in R^{1\times N}$ denote a vector with $N$ features and $y_{i}\in \{-1,1\}$ is label corresponding to $x_{i}$. Let $\vec{1}=(1,1,\cdots,1)$ and $y=(y_{1},y_{2},\cdots,y_{M})$. Define $\phi$ as the mapping function, the original input space can be mapped into the feature space and the inner product $x_{i}x_{j}^{T}$ can be represented as $\phi(x_{i})\phi(x_{j})^{T}$ in this feature space. Let $I_{M\times M}$ represents the $M\times M$ identity matrix and $e_{i}$ denotes the i-th row of $I_{M\times M}$.  Define the matrix of containing original data $A=(x_{1}^{T},x_{2}^{T},\cdots,x_{M}^{T})^{T}$ and $\phi(A)=(\phi(x_{1}^{T}),\phi(x_{2}^{T}),\cdots,\phi(x_{M}^{T})^{T})$. Let the kernel matrix $K$ can be $K_{ij}=K(x_{i},x_{j})=\phi(x_{i})\phi(x_{j})^{T}$. The singular value decomposition (SVD) of matrix $A$ is $A=\sum_{i=1}^{R}\sigma_{i}u_{i}v_{i}^{T}$ where $\sigma_{i}$ and $u_{i},v_{i}$ are eigenvalues and the corresponding eigenvectors respectively. The SVD of $\phi(A)$ can be denoted as $\phi(A)=U_{M\times M}\Sigma_{M\times N} V_{N\times N}^{T}=U_{M\times R}\Sigma_{R\times R} V_{N\times R}^{T}$ where $U_{M\times M}=(U_{M\times R},U_{M\times(M-R)})=(u_{1},u_{2},\cdots,u_{M})$ and $V_{N\times N}^{T}$ has similar expression. The generalized inverse of $\phi(A)$ is $\phi(A)^{\dag}=V_{R\times N} \Sigma_{R\times R}^{-1} U_{M\times R}^{T}$. The distance between the two vectors $x_{i},x_{j}$ can be $|x_{i}-x_{j}|=\sqrt{\sum_{k=1}^{M}(x_{ik}-x_{jk})^{2}}$.\\
\subsection{The SVM and LS-SVM models}
For the large-scale datasets $\{x_{i},y_{i}\},i=1,2,\cdots,M$, the goal of SVM or LS-SVM is to output parameters $w \in R^{N_{\phi}\times1}$ and $b \in R$ where the value of $N_{\phi}$ can be related to $\phi$. The new data $\hat{x}$ can be distinguished by the model containing these parameters: $f(\hat{x})=sgn(w^{T}\phi(\hat{x})^{T}+b)$.\\
\indent The soft margin SVM model finds these optimal parameters $w,b$ by solving the following optimization task:
\begin{align}
&\mathop{min}_{w,b}\frac{1}{2}\|w\|^{2}+\frac{1}{2}\gamma\sum_{i=1}^{M}\xi_{i}\notag\\
s.t.~y_{i}(w^{T}\phi(x_{i})&+b)\geq1-\xi_{i},\xi_{i}\geq0,i=1,2,\cdots,M.
\end{align}
where $\gamma$ is a penalty value, $\xi_{i},i=1,\cdots,M$ are the slack variables. By solving the dual problem of Eqs.(1) with Lagrangian multipliers $\alpha_{i} ,i=1,2,\cdots,M$ where $\alpha=(\alpha_{1},\alpha_{2},\cdots,\alpha_{M})$, parameters $w$ can be represented as $w=\sum_{i=1}^{M}\alpha_{i}x_{i}$. This SVM model satisfies the related KKT condition
\begin{align}
\begin{cases}
&\alpha_{i}(f(x_{i})y_{i}-1+\xi_{i})=0,\\
&1-\xi_{i}-f(x_{i})y_{i}\leq0,\\
&\alpha_{i}\geq0.
\end{cases}
\end{align}
\indent In this KKT condition, there are $f(x_{i})y_{i}=1-\xi_{i}$ and it shows that $x_{i}$ is a support vector if the parameter $\alpha_{i}\neq 0$ . The parameter $\alpha_{i}$ has no effect on $f(x_{i})$ when $\alpha_{i}=0$. The SVM model has sparsity because of these parameters $\alpha_{i}=0$.\\
\indent The LS-SVM model uses equality instead of inequality in terms of constraint condition and finally the quadratic programming (QP) problem can be transformed to solve matrix inversion problem.
\begin{align}
&\mathop{min}_{w,b}\frac{1}{2}\|w\|^{2}+\frac{1}{2}\gamma\sum_{i=1}^{M}\xi_{i}\notag\\
&s.t.~y_{i}(w^{T}\phi(x_{i})+b)=1-\xi_{i},\xi_{i}\geq0.
\end{align}
where this equality can be replaced as $(w^{T}\phi(x_{i})+b)=y_{i}-y_{i}\xi_{i}$. The Eqs.(3) can be used to construct Lagrange function with Lagrangian multipliers $\alpha_{i}$ and other parameters. By taking partial derivatives of this function and eliminating other parameters, $\alpha,b$ satisfy $w^{T}=\sum_{i=1}^{M}\alpha_{i}\phi(x_{i})^{T}$ and have the following formulation
\begin{equation}
\left[\begin{array}{c}
  \alpha \\
  b
\end{array}\right]
=\left[
            \begin{array}{cc}
              K+\gamma^{-1}I_{MM} & \vec{1}^{T} \\
              \vec{1} & 0\\
            \end{array}
          \right]^{-1}
          \left[\begin{array}{c}
                        y \\
                        0
                      \end{array}\right]
          =M_{s}^{-1}y_{s}.
\end{equation}
\indent The original LS-SVM model corresponding to Eqs.(4) is inappropriate for calculating a single entry of vectors and we employ the equivalent model (PLS-SVM)~\cite{zhou2015sparse} instead of the original one. PLS-SVM can be obtained via the representer theorem~\cite{scholkopf2001generalized} and the corresponding optimization problem can be described as follows:
\begin{equation}
\mathop{min}_{\alpha,b}\frac{1}{2}\alpha(\gamma^{-1}K+KK^{T})\alpha^{T}+\alpha K(\vec{1}^{T}b-y)-y^{T}\vec{1}^{T}b+\frac{Mb^{2}}{2}.
\end{equation}
\indent The Eqs.(4) can be changed to the following formulation:
\begin{align}
(\gamma^{-1}K+KK^{T}&-\frac{1}{M}K\vec{1}^{T}\vec{1}K)\alpha^{T}=K(y^{T}-\frac{\vec{1}y^{T}}{M}\vec{1}^{T}).\notag\\
&b=\frac{1}{M}(\vec{1}y-\vec{1}K\alpha^{T}).
\end{align}
\subsection{Quantum matrix inversion assisted with continuous variables}
\indent Based on the Ref.\cite{zhang2019realizing,arrazola2018quantum}, we introduce an improved quantum hybrid variables method as this operation on matrix inversion. Compared with original method [32], we can simply the algorithm. Moreover, the width $L$ can be employed to control size of eigenvalues. Assume the SVD composition of $M_{s}$ is $M_{s}=\sum_{i=1}^{M+1}\hat{\lambda}_{i}\hat{u}_{i}\hat{u^{\dag}}_{i}^{T}$. The inverse matrix $M_{s}^{-1}$ has following integral form:
\begin{equation}
      M_{s}^{-1}=\frac{i}{\sqrt{2\pi}}\int_{-\infty}^{\infty}\int_{-\infty}^{\infty}\psi(p_{1})p_{2}e^{-\frac{p_{2}^{2}}{2}}e^{iM_{s} p_{1}p_{2}}dp_{1}dp_{2}.
\end{equation}
where
\begin{equation}
      \psi(p_{1})=\{
            \begin{array}{cc}
              1, & 0<p_{1}\\
              0, & otherwise\\
            \end{array}.\notag
\end{equation}
\indent Then our goal is to use continuous quantum variables to perform matrix inversion. We use CV quantum states
\begin{align}
      |q\rangle&=\int_{-\infty}^{\infty}\psi(p_{1})|p_{1}\rangle dp_{1} , \notag\\ |g\rangle&=\frac{i}{\sqrt{2\pi}}\int_{-\infty}^{\infty}p_{2}e^{-\frac{p_{2}^{2}}{2}}|p_{2}\rangle dp_{2}.
\end{align}
where the step function state of Eqs.(7) is unphysical and the finite width $L$ can be used to approximate ideal step function
\begin{equation}
      \psi(p_{1})=\{
            \begin{array}{cc}
              1, & 0<p_{1}<L\\
              0, & otherwise\\
            \end{array}.\notag
\end{equation}

\indent The quantum form of the operation $M_{s}^{-1}u_{i}$ is equal to this process: constructing the unitary operation $e^{iM_{s}\hat{p_{1}}\hat{p_{2}}}$ [14,26] and performing this operation on $|\hat{u}_{i}\rangle|q\rangle|g\rangle$, we approximately obtain the $\hat{\lambda}_{i}^{-1}|\hat{u}_{i}\rangle$ with a certain probability after using homodyne dectection.
\section{\label{sec:level3}HVQ-SVM and QSLS-SVM algorithms}
In this section we design LS-SVM algorithm of quantum version and this algorithm consists of HVQ-SVM and QSLS-SVM algorithms. The HVQ-SVM algorithm is a quantum algorithm in overall procedure and can be better than all-qubits version. The QSLS-SVM algorithm can be employed to obtain a classical solution with equivalent parameters and consists of quantum part and classical part.
\subsection{The HVQ-SVM Algorithm}
We first transform this LS-SVM model to a quantum problem. The kernel classification function of LS-SVM model have the following form:
\begin{align}
      f(\hat{x})&=sgn(w^{T}\phi(\hat{x})^{T}+b)=sgn(\phi(\hat{x})\phi(A)^{T}\alpha^{T}+b) \notag\\
                &=sgn((\phi(\hat{x}),1)\left[
                                     \begin{array}{cc}
                                       \phi(A)^{T} & 0 \\
                                       0 & 1 \\
                                     \end{array}
                                   \right]\left[\begin{array}{c}
                                            \alpha \\
                                            b
                                          \end{array}\right])\notag\\
                &=sgn(x_{s}A_{s}\alpha_{s}).
\end{align}
\indent Then the vectors $x_{s}$ and $y_{s}$ are encoded as quantum states $|\psi_{x_{s}}\rangle$ and $|\psi_{y_{s}}\rangle$ respectively via QRAM~\cite{giovannetti2008quantum,giovannetti2008architectures}. And the matrices $A_{s}$ can be loaded as $|\psi_{A_{s}}\rangle$. Thus, the classification problem can be inner product of the two quantum states $|\psi_{A_{s}}\rangle$ and $|\psi_{x_{s}}\rangle M_{s}^{-1}|\psi_{y_{s}}\rangle$. The core of this problem is to obtain $M_{s}^{-1}|\psi_{y_{s}}\rangle$. Specifically, the state $|\psi_{y_{s}}\rangle$ has the form $|\psi_{y_{s}}\rangle=\sum_{i=1}^{M+1}y_{i}|i\rangle=\sum_{i=1}^{M+1}c_{i}|\hat{u}_{i}\rangle$ where $y_{M+1}=0$ and the latter is a representation of $|\psi_{y_{s}}\rangle$ on this basis $\{|\hat{u}_{i}\rangle\}$. And $|\psi_{x_{s}}\rangle,|\psi_{A_{s}}\rangle$ have similar expression when we deal with a linear LS-SVM model. Here we take polynomial kernel function $K(x,z)=(\langle x,z\rangle+1)^{d}$ as a example to describe the preparation of $|\psi_{x_{s}}\rangle,|\psi_{A_{s}}\rangle$ when we handle the non-linear LS-SVM model. Assume $\hat{x}'=(\hat{x},1,1),x_{i}'=(x_{i},1,0)^{T},$ and $x_{M+1}'=(0,0,\cdots,0,1)^{T}$, these vectors are encoded as quantum states $|\hat{x}'\rangle,|x_{i}'\rangle,i=1,\cdots,M+1$ and $K(x_{i},x_{j})=\langle x_{i}'|x_{j}'\rangle^{d}$. Then the states $|\psi_{x_{s}}\rangle,|\psi_{A_{s}}\rangle$ can be denoted as $|\psi_{x_{s}}\rangle=\underbrace{|\hat{x}'\rangle\otimes\cdots\otimes|\hat{x}'\rangle}_{d}$ and $|\psi_{A_{s}}\rangle=\sum_{i=1}^{M+1}|i\rangle|\phi(x_{i}')\rangle$ where $|\phi(x_{i}')\rangle=\underbrace{|x_{i}'\rangle\otimes\cdots\otimes|x_{i}'\rangle}_{d}$. The detail description can be found in appendix A.\\
\indent Here we use the hybrid variables method in our algorithm. However, matrix $M_{s}$ is full rank and unitary operation $e^{iM_{s}\hat{p_{1}}\hat{p_{2}}}$ can not be constructed. Matrix inversion can be constructed as a combination of unitary operators~\cite{rebentrost2014quantum}. Therefore, $e^{iM_{s}\hat{p_{1}}\hat{p_{2}}}=\prod_{k=1}^{3}e^{iG_{k}\hat{p_{1}}\hat{p_{2}}}$ where
\begin{equation}
         G_{1}=\left[
            \begin{array}{cc}
              K & 0 \\
              0 & 0\\
            \end{array}
          \right]
         G_{2}=\left[
            \begin{array}{cc}
              \gamma^{-1}I & 0 \\
              0 & 0\\
            \end{array}
          \right]
         G_{3}=\left[
            \begin{array}{cc}
              0 & \vec{1}^{T} \\
              \vec{1} & 0\\
            \end{array}
          \right].
\end{equation}
\indent The whole process of HVQ-SVM Algorithm is described as follows:\\
\indent 1: Initialization. The discrete quantum state is encoded as $|\psi_{y_{s}}\rangle$. Two-mode is initialized in $|q\rangle|g\rangle$. Then the initial quantum state is prepared as $|\psi_{1}\rangle=|\psi_{y_{s}}\rangle|q\rangle|g\rangle$.\\
\indent 2: Phase estimation. Applying the uniform operation $e^{iM_{s}\hat{p_{1}}\hat{p_{2}}}$ on $|\psi_{1}\rangle$ which leads to
\begin{equation}
      |\psi_{2}\rangle=\sum_{k=1}^{M+1}c_{k}|\hat{u}_{k}\rangle e^{i\hat{\lambda}_{k}\hat{p_{1}}\hat{p_{2}}}|q\rangle|g\rangle.
\end{equation}
\indent 3: Homodyne detection. The operation $I\otimes|0_{p_{1}}\rangle\langle0_{p_{1}}|\otimes|0_{p_{2}}\rangle\langle0_{p_{2}}|$ can be applied on $|\psi_{2}\rangle$ and two-mode can be eliminated to obtain:
\begin{equation}
      |\psi_{3}\rangle=\sum_{k=1}^{M+1}c_{k}\lambda_{k}^{-1}|\hat{u}_{k}\rangle.
\end{equation}
where $\langle q|0_{p_{1}}\rangle=\langle r|0_{p_{2}}\rangle=1$. It is equal to perform a homodyne measurement on two-mode and post-select the result $p = 0$.\\
\indent 4: Measurement. Assume $|\psi_{4}\rangle=|\psi_{x_{s}}\rangle|\psi_{3}\rangle$, an ancillary qubit can be used to construct an entangled state $|\psi_{5}\rangle=\frac{1}{\sqrt{2}}(|0\rangle|\psi_{4}\rangle+|1\rangle|\psi_{A_{s}}\rangle)$. Performing a $\sigma_{x}$ measurement and the inner product $\langle A_{s}|\psi_{4}\rangle$ can be obtained.\\
\indent We omit the normalization factor in overall procedure of this algorithm and choose the suitable width $L$ to preserve the fidelity and reduce the error. Notably this error can be taken as a regulation method (see the appendix B).\\
\chapter{\emph{Analysis.}}
We briefly discuss the time complexity of our algorithm. The main part of complexity is the preparation of initial quantum state and construction of unitary transformation $e^{iM_{s}\hat{p_{1}}\hat{p_{2}}}$ or $e^{iG_{k}\hat{p_{1}}\hat{p_{2}}},k=1,2,3$. The initial quantum state can be prepared via qRAM and costs $O(logMN)$. In the quantum phase estimation, the unitary operation $e^{iM_{s}\hat{p_{1}}\hat{p_{2}}}$ can be constructed by this method outlined in Ref.\cite{lau2017quantum,zhang2019realizing}. It costs time $O(\varepsilon^{-1}log(MN))$ where $\varepsilon^{-1}$ denotes precision of generating the eigenvalues.
The total evolution time complexity of our algorithm is $O(\varepsilon^{-1}log(MN))$.
\subsection{The QSLS-SVM algorithm}
\indent There are a series of algorithms~\cite{suykens2000sparse,ling2014towards,mall2015very,silva2015novel,zhou2017new,chen2018sparse} to solve sparsity of LS-SVM. The sparsity can reflect advantage that we can use small samples to reveal the primary performance of entire datasets. Another advantage is that these important samples can be obtained by using quantum algorithm in $O(log[M])$ time. Notably, the quantum algorithm can be used as a subroutine for overall process of algorithm since we obtain classical information by measuring the output result of quantum algorithm. \\
\indent As a better way, the P-LSSVM model can be constructed to solve sparse problem. The method about ny$\ddot{o}$trom approximation~\cite{fowlkes2004spectral,li2014large} outlined in Ref.\cite{suykens2000sparse} can be used to solve this problem. However, the cost of finding a full-rank submatrix of $K$ is expensive. Here we employ the optimal matrix approximation instead of above method to achieve the sparse solution of P-LSSVM.\\
\indent Define the left eigenvector matrix of $\phi(A)$ as $\Omega=U_{M\times M}\in R^{M\times M}$, we show that $\phi(A)_{\Omega}$ and $\phi(A)$ have same result in classification function where $\phi(A)_{\Omega}=\Omega^{T}\phi(A)$. Assume $\alpha_{\Omega}^{T}=\Omega^{T}\alpha^{T},y_{\Omega}^{T}=\Omega^{T}y^{T},\vec{1}_{\Omega}^{T}=\Omega^{T}\vec{1}^{T}$, $K_{\Omega}=\phi(A)_{\Omega}\phi(A)_{\Omega}^{T}$ and the classification function has the following form:
\begin{equation}
f_{\Omega}(\hat{x})=sgn(\phi(\hat{x})\phi(A)_{\Omega}^{T}\alpha_{\Omega}^{T}+b')
\end{equation}
\indent Compared with the Eqs.(9), we have $b=b'$ and $f_{\phi}(\hat{x})=f(\hat{x})$. Let's take it a step further. In Eqs.(13), the inner product $\phi(A)_{\Omega}^{T}\alpha_{\Omega}^{T}$ can be expressed as follows: $\phi(A)_{\Omega}^{T}(\alpha_{\Omega}^{'})^{T}$ where the sparse vector $\alpha_{\Omega}^{'}=(\alpha_{\Omega_{R}},0)$ and vector $\alpha_{\Omega_{R}}$ represents the top-R entries of $\alpha_{\Omega}$. Thus, $\alpha_{\Omega_{R}}$ and $b$ are the valid parameter that we get.\\
\indent In the same way, we denote the top-R entries of $\vec{1}_{\Omega},y_{\Omega}$ as $\vec{1}_{\Omega_{R}},y_{\Omega_{R}}$, respectively. And suppose the submatrix of $K_{\Omega}$ as $K_{\Omega_{R}}=\Sigma_{R\times R}$, the solution Eqs.(6) of P-LSSVM can be simplified as:
\begin{align}
(\gamma^{-1}K_{\Omega_{R}}&+K_{\Omega_{R}}^{2}-\frac{1}{M}K_{\Omega_{R}}\vec{1}_{\Omega_{R}}^{T}\vec{1}_{\Omega_{R}}K_{\Omega_{R}})\alpha_{\Omega_{R}}^{T}\notag\\
&=K_{\Omega_{R}}(y_{\Omega_{R}}^{T}-\frac{\vec{1}y^{T}}{M}\vec{1}_{\Omega_{R}}^{T}),\notag\\
&b=\frac{1}{M}(\vec{1}y^{T}-\vec{1}_{\Omega_{R}}K_{\Omega_{R}}\alpha_{\Omega_{R}}^{T}).
\end{align}
\indent From this formulation, we come to the conclusion that the product $\vec{1}y^{T}$ and the vectors $\vec{1}_{\Omega_{R}},K_{\Omega_{R}},y_{\Omega_{R}}$ are the core of target parameters $\alpha_{\Omega_{R}},b$. Firstly, we encode the vector $y^{T}$ as a quantum state $|\psi_{y}\rangle=\sum_{i=1}^{M}y_{i}|i\rangle$ and apply Hadamard gate $H^{\otimes^{\lceil logM\rceil+1}}$ on $|0\rangle$ to obtain uniform superposition state $|\psi_{\vec{1}}\rangle=\sum_{i=1}^{M}|i\rangle$. The value $\vec{1}y^{T}$ can be calculated by $\langle\psi_{y}|\psi_{\vec{1}}\rangle$. Next, we use the quantum toolbox such as QPCA or QSVT~\cite{lloyd2014quantum,duan2018efficient} to extract singular values $\sigma_{i}$ and  eigenvectors $u_{i},v_{i}$. These singular vectors can be used to calculate $\vec{1}_{\Omega_{R}},y_{\Omega_{R}}$ via the product $u_{i}^{T}y^{T},u_{i}^{T}\vec{1}^{T}$ in quantum form. In addition, to avoid the complex classical calculation caused by large-scale features, e.g. $N=O(M)$, we need to obtain $\phi(\hat{x})\phi(A)_{\Omega}^{T}$ of Eqs.(13) by computing the product $\sigma_{i}\phi(\hat{x})v_{i}$. \\
\indent Based on the above analysis, we divide QSLS-SVM into quantum and classical subroutines. In the quantum part of the algorithm, the goal we want to achieve is to gain the $\vec{1}_{\Omega_{R}},K_{\Omega_{R}},y_{\Omega_{R}}$ and $\phi(\hat{x})\phi(A)_{\Omega}^{T}$. And these information can be employed to solve the Eqs.(13) and Eqs.(14) via classical calculation.\\
\indent There are a couple of other things to notice before we go into the algorithm. The matrix $\phi(A)$ can be loaded as the quantum state $|\psi_{\phi(A)}\rangle=\sum_{i=1}^{M}\sum_{j=1}^{N_{\phi}}\phi(A)_{ij}|i\rangle|j\rangle=\sum_{i=1}^{R}\sigma_{i}|u_{i}\rangle|v_{i}\rangle$ via qRAM~\cite{giovannetti2008quantum,giovannetti2008architectures} and Gram-Schmidt decomposition. And the multiple unitary operations $e^{i\eta K\hat{p}_{3}}$~\cite{lau2017quantum,zhang2019realizing} can be constructed in quantum phase estimation of this algorithm. In this paper, we take QPCA algorithm as the tool, which is based on a single continuous variable~\cite{lloyd1999quantum,lau2017quantum} corresponding to orthogonal conjugate operator $\hat{p}_{3},\hat{q}_{3}$, i.e. $[\hat{p}_{3},\hat{q}_{3}]=i$ and $|q_{3}\rangle_{q_{3}}=\frac{i}{\sqrt{2\pi}}\int dp_{3}e^{-iq_{3}p_{3}}|p_{3}\rangle_{p_{3}}$. The algorithm process is described as follows:\\
\indent 1:State preparation. The training data $\{x_{i}\}$ can be encoded as the quantum state $|\psi_{\phi(A)}\rangle=\sum_{i=1}^{M}\sum_{j=1}^{N_{\phi}}\phi(A)_{ij}|i\rangle|j\rangle$ with kernel map $\phi$. And one mode can be initialized in $|0\rangle_{q_{3}}$. The initial state can be set to $|\psi_{1}^{'}\rangle=|\psi_{\phi(A)}\rangle|0\rangle_{q_{3}}$.\\
\indent 2:Phase estimation. Perform the operation $e^{i\eta K\hat{p}_{3}}$ on $|\psi_{1}^{'}\rangle$ which can lead to
\begin{equation}
      |\psi_{2}^{'}\rangle=\sum_{k=1}^{R}\sigma_{k}|u_{k}\rangle|v_{k}\rangle|\eta\lambda_{k}\rangle_{q_{3}},
\end{equation}
where $\lambda_{k}=\sigma_{k}^{2}$.\\
\indent 3:Homodyne detection. After a homodyne detection on mode $q_{3}$, the state $|\psi_{2}^{'}\rangle$ collapses to $|\psi_{3}^{'}\rangle=|u_{k}\rangle|v_{k}\rangle$ and we obtain the eigenvalue $\lambda_{k}$.\\
\indent 4:Numerical calculation. Vectors $y,\vec{1}$ and new data $\hat{x}$ can be loaded as quantum states $|\psi_{y}\rangle,|\psi_{\vec{1}}\rangle,|\psi_{\phi(\hat{x})}\rangle$, respectively. For the k-th entries of vectors $\vec{1}_{\Omega_{R}},y_{\Omega_{R}}$ and $\phi(\hat{x})\phi(A)_{\Omega}^{T}$, an ancillary qubit can be used to construct entangled states
\begin{align}
&|\beta_{1}\rangle=\frac{1}{\sqrt{2}}(|0\rangle|u_{k}\rangle+|1\rangle|\psi_{\vec{1}}\rangle),\notag\\
&|\beta_{2}\rangle=\frac{1}{\sqrt{2}}(|0\rangle|u_{k}\rangle+|1\rangle|\psi_{y}\rangle),\notag\\
&|\beta_{3}\rangle=\frac{1}{\sqrt{2}}(|0\rangle|v_{k}\rangle+|1\rangle|\psi_{\phi(\hat{x})}\rangle).
\end{align}
\indent Then, perform a $\sigma_{x}$ measurement on the ancillary qubit of the states $|\beta_{j}\rangle,j=1,2,3$ which can obtain the k-th entries of target vectors.\\
\indent 5: Repeat step1 through step4 until we have all the components of the target vectors. An entangled state $|\beta_{4}\rangle=\frac{1}{\sqrt{2}}(|0\rangle|y\rangle+|1\rangle|\psi_{\phi(\hat{x})}\rangle)$ can be constructed to obtain $\vec{1}y^{T}$.\\
\indent 6: Classical calculation. These vectors and $\vec{1}y^{T}$ can be used to get the $\alpha_{\Omega_{R}}$ and $b$ by solving the Eqs.(14). \\
\indent Finally, we identify categories which the new data $\hat{x}$ belongs to via classification function $f(\hat{x})$. However, the new data can not be used alone in the above-presumed large-scale features. Taking small-scale features into account, e.g. $N\ll M$, we can construct classical model after obtaining the target parameters $\alpha_{\Omega_{R}}$ and $b$, i.e. suppose $w_{\Omega}=\phi(A)_{\Omega}^{T}(\alpha_{\Omega}^{'})^{T}$, we have $f(\hat{x})=sgn(w_{\Omega}^{T}\phi(\hat{x})^{T}+b)$. This case is more suitable for linear LS-SVM.\\
\chapter{\emph{Analysis.}}
We briefly discuss the time complexity of our algorithm. The main part of complexity is the preparation of initial quantum state and operation $e^{i\eta K\hat{p}_{3}}$. Similar to the HVQ-SVM algorithm, the total evolution time complexity of the QSLS-SVM algorithm is $O(\varepsilon^{-1}log(MN))$ in step1-ster4. And actually, we need to repeat $O(R)$ times to achieve numerical calculation. In addition, we should spend $O(N+R^{3})$ to for the classical calculation. The total time complexity is $O(N+R^{3}+R(\varepsilon^{-1}log(MN)))$.
\section{\label{sec:level4}DISCUSSIONS AND CONCLUSIONS}
Both HVQ-SVM and QSLS-SVM algorithms have their own advantages and different applications. The former is simplified in algorithm and the latter can get a sparse solution. For the analysis of HVQ-SVM, please refer to appendix B. Here we mainly analyse the QSLS-SVM algorithm in this section. In practice, the cost of obtaining all the eigenvalues by QPCA algorithm is expensive, and we discuss this problem in the following paragraph. In addition, we also analyze the effect of our method on SVM.\\
\indent In the QSLS-SVM algorithm mentioned above, the requirement for the matrix $\phi(A)$ is low-rank and we only extract large eigenvalues, i.e. we may get top-T eigenvalues where $T\leq R$. Assume $f(\hat{x})=sgn(g(\hat{x})),f_{\Omega}(\hat{x})=sgn(g_{\Omega}(\hat{x}))$, there exists the error between $g(\hat{x})$ and $g_{\Omega}(\hat{x})$. Therefore, the selection of eigenvalues is crucial (see the appendix C). For this case, we should use the QSVT algorithm and its expansion~\cite{duan2018efficient,Duan2019Quantum,lin2019improved} which can employ threshold to eliminate small eigenvalues. Our algorithm is also suitable for high-rank matrix by running an improved QPCA algorithm~\cite{lin2019improved} which is based on threshold. Besides, the quantum-inspired algorithm~\cite{gilyen2018quantum,tang2018quantum} can handle the low-rank matrices which is helpful to get the target eigenvalues.\\
\indent The goal of QSLS-SVM algorithm is mainly to obtain the $\alpha_{\Omega}^{'}$ instead of $\alpha$. This idea may also play an important role in the dual form solution of support vector machines (see appendix D). Obviously, the parameters of our algorithm are one of the solutions of SVM.
\indent In summary, we have investigated quantum-enhanced least-square SVM with two quantum algorithms:  a simplified quantum algorithm for least-square SVM assisted with continuous variables, and a hybrid quantum-classical procedure that allows sparse solutions for least-square SVM with quantum-enhanced feature maps. The algorithmic complexity of the two algorithms costs $O(\varepsilon^{-1}log(MN))$ and $O(N+R^{3}+R(\varepsilon^{-1}log(MN)))$, respectively, and both give exponential speed-up over the sample size M. The algorithms proposed here may be integrated to tackle classically difficult machine learning tasks with classical intractable quantum feature map.

\appendix
\section{The construction of kernel quantum state: Guass and polynomial kernel functions}
In this section we discuss the kernel map in Hilbert space. And in this paper we only describe the polynomial kernel function $K(x,z)=(\langle x|z\rangle+1)^{d}$ and radial basis function $K(x,z)=e^{-|x-z|^{2}/2\omega^{2}}$. For $K(x,z)=(\langle x,z\rangle+1)^{d}$, we take $d=2$ and $x=(x_{1},x_{2}),z=(z_{1},z_{2})$ as a example to analyze this situation. Then we have
\begin{align}
      K(x,z)&=x_{1}^{2}z_{1}^{2}+2x_{1}x_{2}z_{1}z_{2}+x_{2}^{2}z_{2}^{2}+2x_{1}z_{1}+2x_{2}z_{2}+1\notag\\
            &=\phi(x)\phi(z)^{T}.
\end{align}
where $\phi(x)=(x_{1}^{2},x_{1}x_{2},x_{1}x_{2},x_{2}^{2},x_{1},x_{2},x_{1},x_{2},1)$ and $\phi(z)$ has similar form. Assume the vector $(x,1)$ can be encoded the $|x\rangle=x_{1}|00\rangle+x_{2}|01\rangle+|10\rangle=\sum_{i=1}^{3}x_{i}|i\rangle$, we have $|\phi(x)\rangle=\sum_{i,j=1}^{3}x_{i}x_{j}|i\rangle|j\rangle$. Finally, we consider a general case. Let the data $x=(x_{1},\cdots,x_{M}),x'=(x,1)$ and $|\phi(x)\rangle=\underbrace{|x'\rangle\cdots|x'\rangle}_{d}=\sum_{k_{1},\cdots,k_{d}=1}^{M+1}x_{k_{1}}\cdots x_{k_{d}}|k_{1}\rangle\cdots|k_{d}\rangle$. Of course, the entries of vector $\phi(x)$ change the order.
\begin{equation}
      K(x,z)=\langle\phi(x)|\phi(z)\rangle=\langle x'|z'\rangle^{d}.
\end{equation}
\indent For radial basis function $K(x,z)=e^{-|x-z|^{2}/2\omega^{2}}$, we consider the $K(x,z)=e^{-|x-z|^{2}/2\omega^{2}}=e^{-\sum_{i=1}^{M}(x_{i}-z_{i})^{2}/2\omega^{2}}$. Then we have $K(x,z)=e^{\sum_{i=1}^{M}-x_{i}^{2}/2\omega^{2}}e^{\sum_{i=1}^{M}x_{i}z_{i}/\omega^{2}}e^{\sum_{i=1}^{M}-z_{i}^{2}/2\omega^{2}}$. Inspired by Ref.\cite{rahimi2008random,talwalkar2010matrix,ring2016approximation}, we employ the limitation to solve this separation problem of $h(x,z)=e^{\sum_{i=1}^{M}x_{i}z_{i}/\omega^{2}}=e^{xz^{T}/\omega^{2}}$ which can be denoted as $h(x,z)=\phi(x)\phi(z)^{T}$. The limitation has the following form:
\begin{equation}
      \mathop{lim}_{k\rightarrow \infty}(1+\frac{s}{k})^{k}=e^{s}.
\end{equation}
\indent Thus, there exist a constant $N_{0}$ and accuracy $\delta$ such that $|(1+\frac{s}{k})^{k}-e^{s}|<\delta$ if the integer $k$ satisfies $k>N_{0}$. The function $h(x,z)$ can be approximated as $(1+\frac{xz^{T}/\omega^{2}}{k})^{k}$ with accuracy $\delta$. So we can transform separation problem of $h(x,z)$ to separation problem of $c(x,z)=1+\frac{xz^{T}/\omega^{2}}{k}$. Furthermore, we consider function $c'(x,z)=xz^{T}+\omega^{2}k$. Obviously, $c'(x,z)$ can be denoted as $c'(x,z)=(x,\omega\sqrt{k})(z,\omega\sqrt{k})^{T}$. The vector $x'=(x,\omega\sqrt{k})$ can be loaded as a quantum state $|x'\rangle$. The final state can be denoted as tensor product of this $k$ quantum states, e.g. $|\phi(x')\rangle=\underbrace{|x'\rangle\otimes\cdots\otimes|x'\rangle}_{k}$. Taking normalization coefficient into account, we have
\begin{equation}
      |\phi(x')\rangle=\frac{1}{\sqrt{N_{x'}}}\sum_{i_{1},\cdots,i_{k}=1}^{M+1}\frac{x_{i_{1}}\cdots x_{i_{k}}}{(\omega\sqrt{k})^{-k}}|i_{1}\rangle\cdots|i_{k}\rangle.
\end{equation}
where $N_{x'}^{-1}=(\frac{xx^{T}+\omega\sqrt{k}}{(\omega\sqrt{k})})^{-k}$ is the normalization factor and satisfies $\mathop{lim}_{k\rightarrow \infty}N_{x'}^{-1}=e^{\sum_{i=1}^{M}-x_{i}^{2}/\omega^{2}}$. Eventually, we obtain a quantum state about approximating radial basis function with accuracy $\delta$ and the preparation of Eqs.(A.4) costs time $O(N_{0}log(MN))$. More broadly, the other kernel functions such as exponential and rational quadratic kernel functions can also achieve this expression such as Eqs.(A.4).\\
\indent However, if we encode $\phi(A)$ as a quantum state $|\phi(A)\rangle=\sum_{i=1}^{M}|\phi(x_{i})||i\rangle|\phi(x_{i})\rangle$, there actually is $|\phi(A)\rangle\propto\phi(A)/e^{\sum_{i=1}^{M}-x_{i}^{2}/2\omega^{2}}$ and this affects our result. In this paragraph, we give instructions. Assume the matrix $\varphi_{A}$ is
\begin{equation}
\varphi_{A}=\left[
  \begin{array}{cccc}
    x_{1}' & 0 & \cdots & 0 \\
    0 & x_{2}' & \cdots & 0 \\
    \vdots & \vdots & \ddots & \vdots \\
    0 & 0 & \cdots & x_{M}' \\
  \end{array}
\right]\in R^{M\times (M+1)^{2}},
\end{equation}
where $x_{i}'=(x_{i},\omega\sqrt{2k}),i=1,2,\cdots,M$. From Eqs.(A.5) we can draw a conclusion that the matrix $\varphi_{A}$ is one-sparse and can be decomposed as
\begin{align}
\varphi_{A}&=I_{M\times M}\Sigma_{\varphi_{A}}\left[
  \begin{array}{ccc}
    \frac{x_{1}'}{\sqrt{x_{1}'(x_{1}')^{T}}}  & \cdots & 0 \\
    \vdots  & \ddots & \vdots \\
    0  & \cdots & \frac{x_{M}'}{\sqrt{x_{M}'(x_{M}')^{T}}} \\
  \end{array}
\right]\notag\\
&=I_{M\times M}\Sigma_{\varphi_{A}} F^{\dag}=\sum_{i=1}^{M}\sqrt{x_{i}'(x_{i}')^{T}}e_{i}f_{i}^{\dag}.
\end{align}
where $\Sigma_{\varphi_{A}}=diag({\sqrt{x_{i}'(x_{i}')^{T}}})$ and $e_{i}$ represents the i-th column of $I_{M\times M}$. Then we set $\Psi=\left(
                        \begin{array}{cc}
                          0 & \varphi_{A} \\
                          \varphi_{A}^{T} & 0 \\
                        \end{array}
                      \right)
$, $\Psi$ are one-sparse and can be used to construct the unitary transformations $e^{-i\Psi^{1}t}$. The eigenvectors of $\Psi$ can be $(e_{i}^{T},0)^{T},(0,f_{i}^{T})^{T}$. The $(e_{i}^{T},0)^{T}$ can be mapped into the quantum state $|i\rangle$. According to the Ref.[37], we have
\begin{equation}
     |i\rangle|0\rangle\rightarrow|i\rangle| \sqrt{x_{i}'(x_{i}')^{T}}\rangle\rightarrow\frac{1}{\sqrt{x_{i}'(x_{i}')^{T}}}|i\rangle.
\end{equation}
\indent This step can be repeated $k$ times to get the state $\frac{1}{\sqrt{(x_{i}'(x_{i}')^{T})^{k}}}|i\rangle$, e.g. the target state $(\omega^{2}k)^{-k/2}e^{\sum_{j=1}^{M}-x_{ij}^{2}/2\omega^{2}}|i\rangle$. Thus, the matrix $\phi(A)$ can be prepared as
\begin{align}
     &|\phi(A)\rangle:\frac{1}{\sqrt{N_{\phi}}}\sum_{i=1}^{M}|\phi(x_{i}')||i\rangle|\phi(x_{i}')\rangle\rightarrow\notag\\
&\frac{1}{\sqrt{M}}
     \sum_{i=1}^{M}\sum_{j_{1},\cdots,j_{k}=1}^{M+1}e^{\sum_{j=1}^{M}-x_{ij}^{2}/2\omega^{2}}x_{ij_{1}}\cdots x_{ij_{k}}|i\rangle|j_{1}\rangle\cdots|j_{k}\rangle.
\end{align}
\indent As a matter of fact, the selection of $k$ is a important problem and it costs more expensive if $\langle\phi(x_{p_{1}}),\phi(x_{p_{2}})\rangle>1$. However, normalization coefficient can bring better result. Assume $|\phi(x_{p_{1}})\rangle=\frac{1}{\sqrt{N_{p_{1}}}}\sum_{j}\phi(x_{p_{1}})_{j}|j\rangle$, we have $\langle\phi(x_{p_{1}}),\phi(x_{p_{2}})\rangle=\frac{1}{\sqrt{N_{p_{1}}N_{p_{2}}}}\sum_{j}\phi(x_{p_{1}})_{j}\phi(x_{p_{2}})_{j}$ where $N_{p_{i}}=|x_{p_{i}}|,i=1,2$. Because the inequality $\frac{x+y}{2}\geq\sqrt{xy},x,y\geq0$, we have $\langle\phi(x_{p_{1}}),\phi(x_{p_{2}})\rangle\leq\frac{1}{2}(\frac{1}{N_{p_{1}}}\sum_{j}\phi(x_{p_{1}})_{j}^{2}+\frac{1}{N_{p_{2}}}\sum_{j}\phi(x_{p_{2}})_{j}^{2})\leq1$. Thus, the Eqs.(A.3) has $e^{s}\leq e$ and it shows we can reduce the size of $N_{0}$ to achieve the accuracy $\delta$. Finally, the Eqs.(A.8) can be used to construct the matrix $K$
\begin{equation}
     tr_{2}|\phi(A)\rangle\langle\phi(A)|=\sum_{p_{1},p_{2}=1}^{M}e^{\frac{x_{p_{1}}x_{p_{2}}^{T}}{\sqrt[k]{M}\omega^{2}}}e^{-\frac{x_{p_{1}}x_{p_{1}}^{T}+x_{p_{2}}x_{p_{2}}}{2\omega^{2}}}|p_{1}\rangle\langle p_{2}|.
\end{equation}
\section{Error of HVQ-SVM and Regularzation}
In this section, we discuss the influence of error. The hybrid quantum variables method exists the error since we need to define the width $L$ to preserve precision and the error in homodyne detection. In overall procedure of HVQ-SVM algorithm, we call it the fidelity problem and analyze whether the fidelity problem can bring us good effects.\\
\indent We combine finite squeezing method~\cite{zhang2019realizing} to introduce the fidelity problem. According to the Ref.\cite{zhang2019realizing}, the quantum state using finite squeezing method can be represented as
\begin{equation}
      |\digamma\rangle=\sum_{i}\lambda_{i}B(Q_{1},Q_{2})|\psi_{u_{i}}\rangle|\psi_{v_{i}}\rangle.
\end{equation}
where $B(Q_{1},Q_{2})\sim e^{-(Q_{1}^{2}+Q_{2}^{2})/2\alpha_{i}^{2}s^{2}}/s\alpha_{i}$, two qumodes $(Q_{1},Q_{2})=(p_{1},p_{2})$ is post-selection, $\alpha_{i}=\eta(\lambda_{i}^{2}+\chi)$,$\lambda_{i}$ is the eigenvalue of the matrix.$\eta$ and $\chi$ are both parameters. $B(Q_{1},Q_{2})$ satisfies $\alpha_{i}^{2}s^{4}\sim\varepsilon_{q}^{-1}$ where $\varepsilon_{q}^{-1}$ is a error. What we want to explore is the effect of $B(Q_{1},Q_{2})$ on $\lambda_{i}$. Taking a classical representation into account, we have
\begin{equation}
      A^{+}=\tau\sum_{i}\lambda_{i}B(Q_{1},Q_{2})u_{i}v_{i}.
\end{equation}
where $\tau$ is the product of the normalization factor and $A^{+}$ is equal to $|\digamma\rangle$. Changing one form of expression, the formula is
\begin{equation}
      \lambda_{i}B(Q_{1},Q_{2})=\frac{\lambda_{i}}{\lambda_{i}^{2}+\chi}C(s_{1},\lambda_{i}).
\end{equation}
where
\begin{equation}
      C(s_{1},\lambda_{i})\sim e^{-1/\alpha_{i}^{2}s_{1}^{2}}/s_{1},s_{1}=2s/(Q_{1}^{2}+Q_{2}^{2}).\notag
\end{equation}
Function $C(s_{1},\lambda_{i})$ satisfies the inequality:
\begin{equation}
      1=\frac{C(s_{1},\lambda_{1})}{C(s_{1},\lambda_{1})}\geq\frac{C(s_{1},\lambda_{2})}{C(s_{1},\lambda_{1})}\geq\cdots\geq\frac{C(s_{1},\lambda_{d})}{C(s_{1},\lambda_{1})}.
\end{equation}
Equ.(29) is modified to
\begin{equation}
      \sum_{i}\frac{\lambda_{i}}{\lambda_{i}^{2}+\chi}u_{i}v_{i}=\tau^{'}\sum_{i}\frac{\lambda_{i}}{\lambda_{i}^{2}+\chi}\cdot\frac{C(s_{1},\lambda_{i})}{C(s_{1},\lambda_{1})}u_{i}v_{i}.
\end{equation}
\indent This formula shows that finite squeezing factor regularization is similar to $L_{2}$ regularization, which is a process of reduction along singular value vectors. Our method has same result. Here, we also consider the error of homodyne detection. In HVQ-SVM Algorithm, we can not employ the unitary $U_{p}=I\otimes|0_{p_{1}}\rangle\langle0_{p_{1}}|\otimes|0_{p_{2}}\rangle\langle0_{p_{2}}|$ since it can cause this result that the probability of obtaining target state is 0~\cite{arrazola2018quantum}. Assume $|\varepsilon_{q}\rangle=\frac{1}{\pi^{1/4}\sqrt{\varepsilon_{q}}}$, then $U_{p}=I\otimes|\varepsilon_{q}\rangle\langle\varepsilon_{q}|\otimes|\varepsilon_{q}\rangle\langle\varepsilon_{q}|$ and the Eqs.(12) can be improved as

\begin{equation}
      \sum_{k=1}^{T}c_{k}\lambda_{k}^{-1}\hat{F}(\lambda_{k})|\hat{u}_{k}\rangle.
\end{equation}
where
\begin{equation}
      \hat{F}(\lambda_{k})=\frac{1-e^{-L^{2}(\lambda_{k}^{2}+\varepsilon_{q}^{2}+\varepsilon_{q}^{4})/2(1+\varepsilon_{q}^{2})}}{1+(\varepsilon_{q}^{2}+\varepsilon_{q}^{4})/\lambda_{k}}.
\end{equation}
Then
\begin{equation}
      1=\frac{\hat{F}(\lambda_{1})}{\hat{F}(\lambda_{1})}\geq\frac{\hat{F}(\lambda_{2})}{\hat{F}(\lambda_{1})}
      \geq\cdots\geq\frac{\hat{F}(\lambda_{T})}{\hat{F}(\lambda_{1})}.
\end{equation}
where
\begin{equation}
      \frac{\hat{F}(\lambda_{k})}{\hat{F}(\lambda_{1})}=\frac{1-e^{-L^{2}(\lambda_{k}^{2}+\varepsilon_{q}^{2}+\varepsilon_{q}^{4})/2(1+\varepsilon_{q}^{2})}}{1-e^{-L^{2}(\lambda_{1}^{2}+\varepsilon_{q}^{2}+\varepsilon_{q}^{4})/2(1+\varepsilon_{q}^{2})}}
      \bullet\frac{1+(\varepsilon_{q}^{2}+\varepsilon_{q}^{4})/\lambda_{1}}{1+(\varepsilon_{q}^{2}+\varepsilon_{q}^{4})/\lambda_{k}}.
\end{equation}
\indent Thus, our method has same effect as $L_{2}$ regularization with global phase factor $\hat{F}(\lambda_{1})$ and can cause other effects with different global phase factors $\hat{F}(\lambda_{k})$.
\section{Error for $g(\hat{x})$ and $g_{\Omega}(\hat{x})$ in QSLS-SVM algorithm}
The expensive cost of obtaining the small eigenvalues leads to the error of ideal output and actual output. In this section, we introduce the error analysis. The error of both functions is
\begin{align}
      E_{\Omega}&=|g(\hat{x})-g_{\Omega}(\hat{x})|\notag\\
                &=|\sum_{i=T+1}^{R}(\alpha_{\Omega})_{i}\sigma_{i}[\phi(\hat{x})^{T}V_{i}+\sigma_{i}(\vec{1}_{\Omega})_{i}]|\notag\\
                &\leq\sigma_{T+1}|\sum_{i=T+1}^{R}(\alpha_{\Omega})_{i}[\phi(\hat{x})^{T}V_{i}+\sigma_{i}(\vec{1}_{\Omega})_{i}]|.
\end{align}
\indent For the given training data and kernel map $\phi$, the factors of Eqs.(C.1), except for $T$ and $\hat{x}$, is fixed. It shows that we can adjust the threshold $\tau=\sigma_{T}$ to reduce the error. On the other hand, the new data $\hat{x}$ is a uncontrollable factor which has a effect on $E_{\Omega}$ and this is reflected in some data.
\section{Apply QSLS-SVM method to soft margin SVM}
\indent The dual problem of soft margin SVM model corresponding to Eqs.(5) is
\begin{align}
\mathop{max}_{\alpha}\sum_{i=1}^{M}&\alpha_{i}y_{i}-\frac{1}{2}\sum_{i=1}^{M}\sum_{j=1}^{M}\alpha_{i}\phi(x_{i})\phi(x_{j})^{T}\alpha_{j},\notag\\
&s.t. \sum_{i=1}^{M}\alpha_{i}=0,0\leq\alpha_{i}\leq\frac{1}{2}\gamma.
\end{align}
\indent The Eqs.(17) with the vectors $\alpha_{\Omega}^{T},y_{\Omega}^{T},\vec{1}_{\Omega}^{T},K_{\Omega}$ can be simplified to
\begin{align}
\mathop{max}_{\alpha_{\Omega}}\sum_{i=1}^{R}(\alpha_{\Omega})_{i}&[(y_{\Omega})_{i}-\frac{1}{2}\lambda_{i}(\alpha_{\Omega})_{i}]+\sum_{i=R+1}^{M}(\alpha_{\Omega})_{i}(y_{\Omega})_{i},\notag\\
s.t. &\vec{1}_{\Omega}\alpha_{\Omega}^{T}=0,0\leq\alpha_{\Omega}\leq\frac{1}{2}\gamma\vec{1}_{\Omega}.
\end{align}
\indent The solution of QSLS-SVM algorithm is actually a solution of Eqs.(D.2). And the Eqs.(18) may be helpful for us to solve QP problem. For instance, we can think about ignoring $(\alpha_{\Omega})_{i},i=R+1,\cdots,M$ since these parameters have no effect on the output of SVM.\\
\end{document}